\documentclass[aps,prb,twocolumn,groupedaddress]{revtex4-2}
\usepackage{graphicx} 
\usepackage{bm,amsmath,amssymb,physics,color,ulem}
\newcommand{\ii}{{\mathrm{i}}}
\newcommand{\ee}{{\mathrm{e}}}
\begin{document}
\title{Non-Hermitian skin effect and lasing of absorbing open-boundary modes in photonic crystals}
\author{Tetsuyuki Ochiai}
\affiliation{Research Center for Functional Materials, National Institute for Materials Science (NIMS), Tsukuba 305-0044, Japan}
\date{\today}

\begin{abstract}
We explore absorbing open-boundary modes in non-Hermitian photonic systems. The modes have a continuum spectrum in the infinite-system-size limit and can exhibit the non-Hermitian skin effect. In contrast to the conventional non-Hermitian skin modes under the fixed-end open-boundary condition, the modes concerned  exhibit a strongly size-dependent spectrum that gradually converges to the non-Bloch-band dispersion. The modes correspond to the poles of the $S$ matrix, and are closely related to the lasing. 
We demonstrate these properties in a two-dimensional non-Hermitian  photonic crystal with gain having a point-(pseudo)gap topology. 
\end{abstract}
\maketitle

\section{Introduction}

Recently, much attention has been paid to non-Hermitian systems of quantum particles \cite{PhysRevLett.77.570,moiseyev2011non}, photonics \cite{feng2017non,ota2020active}, mechanics \cite{ghatak2020observation}, and so on.  
The non-Hermiticity generally results in a lifting of the eigenvalues away from the real axis, giving rise to many fascinating phenomena tied with topology \cite{PhysRevX.9.041015,PhysRevLett.124.086801,PhysRevLett.124.056802,RevModPhys.93.015005}. 

Among all, the non-Hermitian skin effect (NHSE) \cite{PhysRevLett.121.086803} is intriguing and is not available in Hermitian systems.  
In the NHSE, the bulk eigenmodes are localized near the boundary, depending on the boundary condition and the conventional bulk-boundary correspondence \cite{hatsugai1993cna} apparently becomes invalid. 
 The physics behind the NHSE is fertile and inspires  applications such as sensing \cite{PhysRevResearch.2.013058}.

So far, the NHSE has been explored mainly in tight-binding models, where the open boundary  with vanishing field components is usually employed. 
The tight-binding model is a discrete model defined on  lattice sites.   
The open boundary condition with vanishing field components is a natural boundary condition in finite lattices. 
These two items implicitly rely on electronic systems, where the electrons are often tightly confined  in atomic orbitals and bounded in media  by work functions.

If we turn our attention to photonic systems, we will find that such a tight-binding description with the  open boundary is not necessarily realistic, although it is widely used in various theoretical treatments. 
The tight-binding model is often employed in coupled cavity systems, and the open boundary condition is represented   
 as the perfect-electric-conductor (PEC) or perfect-magnetic-conductor (PMC) boundary condition.  
However, the radiation field is generally extended in entire photonic systems and nonvanishing near-fields at the boundary often play important roles.

One of the distinctive features of photonic systems is that they interact strongly with outer systems. 
 The outer systems act as reservoirs of the continuum radiation modes \cite{tannoudji1992atom}. As a result, boundaries in photonic systems provide dissipative or decay routes to the outer systems.  
This dissipation and the intrinsic dielectric dispersion with the Kramers-Kronig relation make the photonic  systems lossy and non-Hermitian.  
Moreover, optical gains are easily implemented in photonic systems by the population inversion. 
Thus, the non-Hermiticity is built-in and quite rich in 
photonic systems. 

Because of these features in photonic systems, various engineered non-Hermiticity has been explored in photonic platforms \cite{feng2017non,ota2020active,pan2018photonic}. However, there have been limited studies on the NHSE in photonic systems so far \cite{PhysRevResearch.2.013280,PhysRevB.104.125416,PhysRevB.104.125109,PhysRevResearch.4.023089, PhysRevLett.129.013903,PhysRevApplied.14.064076,FangHuZhouDing+2022+3447+3456}.

Here, we consider yet another aspect of the NHSE in photonic crystals (PhCs) by introducing gains. 
We focus on 
rather uncovered eigenmodes inherent in photonic systems. The modes focused on here are absorbing open-boundary modes that merge with the reservoir of continuous radiation modes in the outer systems. 
Like the conventional fixed-end open-boundary modes, which do not mix with the reservoir, the modes here can exhibit the NHSE.  
The absorbing open-boundary modes correspond to the poles of the $S$ matrices in finite-thickness PhCs. Since the $S$ matrix is like a ratio between  output and input, the poles imply finite outputs for vanishing inputs. 
Thus, the modes  are directly related to the lasing.  Here, we  explore the fundamental properties of the absorbing open-boundary modes 
and their relation to the lasing in detail.

This paper is organized as follows. 
In Sec. II, we present a theoretical formulation via the $S$ matrix for non-Hermitian PhCs and their eigenmode properties. 
In Sec. III, we give the explicit form of the  $S$ matrix of a certain class of two-dimensional (2D) PhCs. 
In Sec. IV, we present numerical results of the  NHSE in a 2D PhC.   
In Sec. V, we present how the absorbing open-boundary modes affect the lasing in the PhC with gain.     
Finally, in Sec. VI, we summarize the results.

\section{$S$ matrix formalism and (non) Bloch-band theory}

\begin{figure}
	\includegraphics[width=0.45\textwidth]{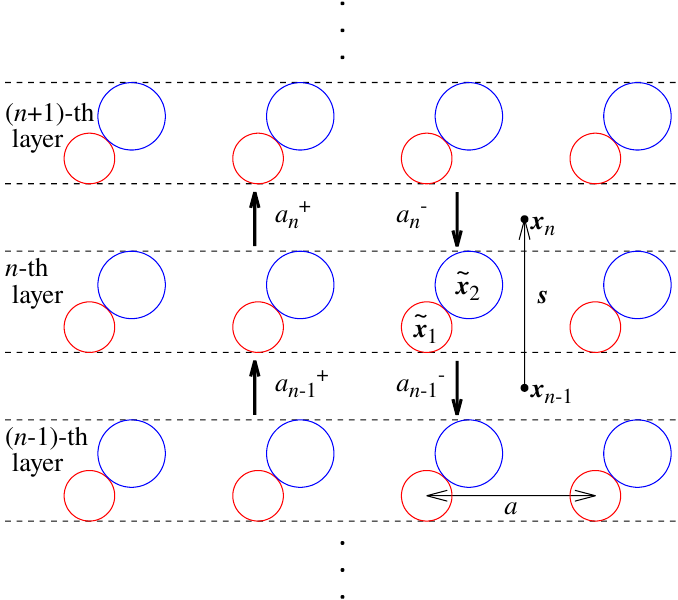}	    
	\caption{\label{Fig_geoPhC} Cross-sectional view of the two-dimensional photonic crystal composed of 
	a parallel array of non-overlapping cylinders.	
	Each layer consists of a one-dimensional periodic array of the cylinders. In the empty region between the $n$-th and $(n+1)$-th layers, the radiation field is expanded by plane waves whose expansion coefficients are symbolically denoted as $a_n^\pm$. 
	}
\end{figure}

In the $S$ matrix formalism, a $D$-dimensional periodic system is  regarded as a stack of ($D-1$)-dimensional periodic ones (labeled by $n$ for the layer index). In between the $n$th and $(n+1)$th layers,  the radiation field is expanded by plane waves whose  
 expansion coefficients are denoted as $a_n^\pm$. Here, the superscript refers to the direction of propagation.  
Figure  \ref{Fig_geoPhC} shows a schematic illustration of a photonic system of $D=2$.  

The $S$ matrix relates input channels to the layer and output channels from the layer. It is defined by  
\begin{align}
&\mqty(a_{n}^+ \\
	a_{n-1}^- )=S(\omega,k_\|)\mqty(a_{n-1}^+ \\
	a_{n}^- ),\label{Eq_defS}\\
&S(\omega,k_\|)=\mqty(S^{++} & S^{+-} \\ S^{-+} & S^{--}).
\end{align}
The $S$ matrix is a function of (complex) angular frequency $\omega$ and Bloch momentum $k_\|$ parallel to the ($D-1$)-dimensional layers. 
The explicit form of the $S$ matrix in a 2D PhC is given in Sec. III.

Once the $S$ matrix is obtained, the transfer matrix is also available. It is defined by 
\begin{align}
&\mqty(a_{n}^+ \\
a_{n}^- )=T(\omega,k_\|)\mqty(a_{n-1}^+ \\
a_{n-1}^- ),\\
&T(\omega,k_\|)=\mqty(T^{++} & T^{+-} \\
T^{-+} & T^{--}),\\
&T^{++}=S^{++}-S^{+-}(S^{--})^{-1}S^{-+},\\
&T^{+-}=S^{+-}(S^{--})^{-1},\\
&T^{-+}=-(S^{--})^{-1}S^{-+},\\
&T^{--} =(S^{--})^{-1}. 
\end{align}
Various properties of non-Hermitian systems, particularly the NHSE,  can be argued in terms of the transfer matrix as shown by Kunst and Dwivedi \cite{PhysRevB.99.245116}. Here, we focus on photonic  aspects of the $S$ matrix and transfer matrix.

The transfer matrix is diagonalized as 
\begin{align}
	&T=U\Lambda U^{-1}, \\
	&U=\mqty( A & B\\ C & D), \quad \Lambda=\mqty(\Lambda_+ & 0\\0&\Lambda_-), \\ &\Lambda_+=\textrm{diag}(\lambda_1,\lambda_2,\dots,\lambda_M), \\ 
	&\Lambda_-=\textrm{diag}(\lambda_{M+1},\lambda_{M+2},\dots,\lambda_{2M}).	
\end{align}
Here, we assume $M$ input (or output) channels, and 
eigenvalues $\lambda_i$ are ordered such that  $|\lambda_1|\ge |\lambda_2|\ge \dots \ge |\lambda_{2M}|$.  
In Hermitian photonic systems with real $\omega$ and $k_\|$, 
half of the eigenvalues are outside the unit circle in the complex plane of $\lambda$. The other half is inside. This classification corresponds to $\Lambda_+$ and  $\Lambda_-$. 
This property  enables us to derive the reflectance of semi-infinite photonic systems in terms of the eigenvectors of the transfer matrix \cite{Botten:N:M:d:A::6404:p046603:2001,PhysRevB.68.155101,ochiai2021bulk}.

We also note that marginal eigenvalues on the unit circle correspond to Bloch-band modes. 
They are expressed as $\lambda_i=\exp(\ii k_\perp d)$ 
for Hermitian systems, where $k_\perp$ coincides with the Bloch momentum perpendicular to the $(D-1)$-dimensional layer and $d$ is the inter-layer distance. In this way, the transfer matrix provides an on-shell photonic-band-calculation scheme. Namely, a set of $k_\perp$ is obtained as a function of  $\omega$ and $k_\|$. 
In contrast, in an ordinary (off-shell) band calculation, a set of $\omega$ is obtained as a function of $(k_\|,k_\perp)$.

Using the transfer matrix, the $S$ matrix of the  $N$-layer system is written as  
\begin{align}
&\mqty(a_{N}^+ \\ a_{0}^-)
=S_N(\omega,k_\|)
\mqty(a_{0}^+ \\a_{N}^- ),\\
&S_N^{++}=T_N^{++}-T_N^{+-}(T_N^{--})^{-1}T_N^{-+},\\
&S_N^{+-}=T_N^{+-}(T_N^{--})^{-1},\\
&S_N^{-+}=-(T_N^{--})^{-1}T_N^{-+},\\
&S_N^{--} =(T_N^{--})^{-1},
\end{align} 
Note that the transfer matrix $T_N$ of the $N$-layer system is simply equal to $T^N$.

The eigenmodes in the $N$-layer system are strongly tied with the $S_N$ matrix and obtained by imposing a boundary condition. Here, we consider three representative boundary conditions in photonic systems.

The first one is the periodic boundary condition, given by 
$a_N^\pm = a_0^\pm$. 
This condition results in the secular equation 
\begin{align}
	(1-S_N)\mqty(a_N^+ \\ a_0^-)=0.  
\end{align}

The second one is the PEC or PMC boundary condition,  which represents a perfect conductor in the vicinity of the boundary surface. 
This is a fixed-end (namely, Dirichlet- or Neumann-type) boundary, and is conventionally called the "open" boundary in non-Hermitian contexts.   
This boundary condition relates $a_{0(N)}^\pm$  as
\begin{align}
	a_{0(N)}^- =L_{0(N)} a_{0(N)}^+,  
\end{align}
with a linear matrix $L_{0(N)}$, 
giving rise to the secular equation 
\begin{align}
	\qty( 1-\mqty( S_N^{+-}L_N & S_N^{++}L_0^{-1}\\
	S_N^{--}L_N & S_N^{-+}L_0^{-1}) )\mqty(a_N^+ \\ a_0^-)=0.  	
\end{align}

The third one is the absorbing boundary condition  we focus on in this paper. This boundary condition assumes  an open boundary, but 
the field component is not vanishing there.  
It merges unidirectionally with the external radiation modes.  
The only requirement is that the radiation field decays exponentially  away from the boundary without any bouncing \footnote{We should note that this boundary condition is not the ordinary absorbing boundary such as the perfect-matched-layer one \cite{berenger1994perfectly}, in which a reflectionless layer with designed optical losses is implemented outside the system concerned.}. 
In Hermitian photonic systems,  the resulting eigenmodes are available outside the light cone, and represent a guided mode propagating parallel to the boundary.  
In non-Hermitian systems, the notion of the light cone loses its meaning, and the absorbing open-boundary modes can generally emerge.   
In this condition, we impose that  solely outgoing waves exist near the PhC boundary, namely, $a_0^+=a_N^-=0$. 
If we divide the $N$-layer PhC into $(N-n)$-layer and $n$-layer ones, with $n$ an arbitrary integer from 1 to $N-1$, we have 
\begin{align}
&\mqty(a_{N}^+ \\
a_{n}^- )=S_{N-n}(\omega,k_\|)\mqty(a_{n}^+ \\
a_{N}^- ),\\	
&\mqty(a_{n}^+ \\
a_{0}^- )=S_{n}(\omega,k_\|)\mqty(a_{0}^+ \\
a_{n}^- ). 
\end{align}
Under the absorbing boundary condition, we obtain the secular equation for $a_n^+$ as 
\begin{align}
(1-S_{n}^{+-}S_{N-n}^{-+})a_n^+=0. 	
\end{align}

The above secular equations can be written in terms of eigenvalues and eigenvectors of the transfer matrix of $M$ input (or output) channels.  
In a special case of $M=1$, the $S$-matrix becomes simply a $2\times 2$ matrix. The secular equations reduce to   
\begin{align}
&\lambda_1^N=1 \quad \textrm{or} \quad \lambda_2^N=1 \quad \textrm{(periodic)},\\	
&\lambda_1^N=\lambda_2^N \quad \textrm{(PEC/PMC)},\label{Eq_PEC}\\
& AD\lambda_2^N=BC\lambda_1^N \quad \textrm{(absorbing)}. 
\end{align} 
In the limit of $N\to\infty$, the first equation has an infinite number of solutions distributed on the curve determined by  $|\lambda_1|=1$ or  $|\lambda_2|=1$ in the complex plane of $\omega$.
This condition corresponds to real $k_\perp$. 
 The second equation reduces, at $N\to\infty$, to $|\lambda_1|=|\lambda_2|$ that corresponds to the non-Bloch-band dispersion  \cite{PhysRevLett.121.086803,PhysRevB.100.035102}. The third one corresponds to the pole of the $S$ matrix and eventually becomes    
$|\lambda_1|=|\lambda_2|$ in the limit of $N\to\infty$. 
Namely, we have
\begin{align}
\lambda_2=\lambda_1 \qty(\frac{BC}{AD})^\frac{1}{N}\ee^{\ii 2\pi\frac{n}{N}}	\quad (n=1,2,\dots,N). 
\end{align}
In the limit of $N\to \infty$, the solutions are densely distributed on the curve defined  by $|\lambda_1|=|\lambda_2|$. 
In contrast to the solutions of Eq. (\ref{Eq_PEC}), at finite $N$, the solutions are systematically deviated from  the curve of $|\lambda_1|=|\lambda_2|$, because of the prefactor $(BC/AD)^{1/N}$.

If there are $M$ input (or output) channels, the secular equations for the periodic and PEC/PMC boundary conditions reduce to the criteria found in the (non-) Bloch-band theory \cite{PhysRevLett.123.066404,PhysRevB.99.201103}. Namely, 
$|\lambda_M|=1$ or $|\lambda_{M+1}|=1$ for the periodic boundary condition and 
$|\lambda_{M}|=|\lambda_{M+1}|$ for the PEC/PMC boundary  condition. 
These criteria define the curves in the complex frequency plane for a given $k_\|$. If these curves do not coincide with each other, the NHSE occurs.

The secular equation for the absorbing  open-boundary modes becomes 
\begin{align}
&\textrm{det}(1-C^{-1}D\Lambda_-^N B^{-1}A\Lambda_+^{-N})=0.
\end{align}
This determinant emerges in the expression of $T_N^{--}$, so that the solutions correspond to the pole of the $S_N$ matrix. 
To have dense solutions in the $N\to\infty$ limit, we need to have 
\begin{align}
&|\lambda_M|=|\lambda_{M+1}|,\\
&(C^{-1}D)_{M1} (B^{-1}A)_{1M} \qty(\frac{\lambda_{M+1}}{\lambda_M})^N=1.
\end{align}
Therefore, the limiting curve of the spectrum of the absorbing open-boundary modes is the same as in the PEC/PMC boundary condition.

\section{$S$ matrix in two-dimensional photonic crystals}
Let us consider a 2D PhC composed of a periodic array of non-overlapping cylinders as an explicit example.  We assume that light is propagating perpendicular to the cylindrical axis (taken to be the $z$ axis).  Thanks to the inversion symmetry concerning the $z$ axis, the radiation field is decoupled into the transverse-electric (TE) and transverse-magnetic (TM) polarization sectors.

Suppose that the PhC is regarded as the stack of identical layers of a 1D periodic array of cylinders as shown in Fig. \ref{Fig_geoPhC}.  
The relative shift between the adjacent layers is denoted as $\vb*{s}$. 
In the empty space between the $n$th and $(n+1)$th layers, the radiation field is expanded by plane waves  as 
\begin{align}
&\psi_n(\vb*{x})=\sum_{g}\qty(
	a_{ng}^+ \ee^{\ii\vb*{K}_{g}^+\cdot(\vb*{x}-\vb*{x}_n)}
	+ a_{ng}^- \ee^{\ii\vb*{K}_{g}^-\cdot(\vb*{x}-\vb*{x}_n)}),\\
&\vb*{K}_g^\pm=(k_x+g)\hat{x}\pm \Gamma_g\hat{y},  \quad \Gamma_g=\sqrt{q^2-(k_x+g)^2},\\
&q=\sqrt{\frac{\omega^2}{c^2}},  	
\end{align} 
where $\psi$ is either $H_z$ (TE polarization) or $E_z$ (TM polarization), $a_{ng}^\pm$ is the plane-wave-expansion (PWE) coefficient of reciprocal lattice $g[=2\pi\textrm{(integer)}/a]$, and $\vb*{x}_n$ is the reference point satisfying $\vb*{x}_{n}-\vb*{x}_{n-1}=\vb*{s}$.   
The square root is chosen such that its imaginary part is always positive.  
Then, the $S$ matrix is defined as Eq. (\ref{Eq_defS}) for column vector $a_n^\pm\equiv (a_{ng_1}^\pm,a_{ng_2}^\pm,\cdots)^t$.

In the layer Korringa-Kohn-Rostoker (KKR) formalism \cite{korringa1947ceb,kohn1954sse,kambe1967theory}, the explicit form of the $S$ matrix is given by \cite{OHTAKA:N::73:p411-413:1979,Ohtaka:U:A::57:p2550-2568:1998,leung1999computation} 
\begin{widetext}
\begin{align}
&S_{gg'}^{++}=\frac{2}{\Gamma_ga}\sum_{\alpha\alpha'll'}{\rm e}^{{\rm i}{\bm K}_g^+\cdot(\vb*{x}_{n}-\tilde{\bm x}_\alpha)}
(-{\rm i})^l {\rm e}^{{\rm i}l\phi({\bm K}_g^+)}T_{(\alpha l)(\alpha' l')}{\rm i}^{l'} {\rm e}^{-{\rm i}l'\phi({\bm K}_{g'}^+)}
{\rm e}^{{\rm i}{\bm K}_{g'}^+\cdot(\tilde{\bm x}_{\alpha'}-\vb*{x}_{n-1})}+\delta_{gg'},\\
&S_{gg'}^{+-}= 
\frac{2}{\Gamma_ga}\sum_{\alpha\alpha'll'}{\rm e}^{{\rm i}{\bm K}_g^+\cdot(\vb*{x}_{n}-\tilde{\bm x}_\alpha)}
(-{\rm i})^l {\rm e}^{{\rm i}l\phi({\bm K}_g^+)}T_{(\alpha l)(\alpha' l')}{\rm i}^{l'} {\rm e}^{-{\rm i}l'\phi({\bm K}_{g'}^-)}
{\rm e}^{{\rm i}{\bm K}_{g'}^-\cdot(\tilde{\bm x}_{\alpha'}-\vb*{x}_{n})},\\
&S_{gg'}^{-+}=
\frac{2}{\Gamma_ga}\sum_{\alpha\alpha'll'}{\rm e}^{{\rm i}{\bm K}_g^-\cdot(\vb*{x}_{n-1}-\tilde{\bm x}_\alpha)}
(-{\rm i})^l {\rm e}^{{\rm i}l\phi({\bm K}_g^-)}T_{(\alpha l)(\alpha' l')}{\rm i}^{l'} {\rm e}^{-{\rm i}l'\phi({\bm K}_{g'}^+)}
{\rm e}^{{\rm i}{\bm K}_{g'}^+\cdot(\tilde{\bm x}_{\alpha'}-\vb*{x}_{n-1})},\\
&S_{gg'}^{--}=\frac{2}{\Gamma_ga}\sum_{\alpha\alpha'll'}{\rm e}^{{\rm i}{\bm K}_g^-\cdot(\vb*{x}_{n-1}-\tilde{\bm x}_\alpha)}
(-{\rm i})^l {\rm e}^{{\rm i}l\phi({\bm K}_g^-)}T_{(\alpha l)(\alpha' l')}{\rm i}^{l'} {\rm e}^{-{\rm i}l'\phi({\bm K}_{g'}^-)}
{\rm e}^{{\rm i}{\bm K}_{g'}^-\cdot(\tilde{\bm x}_{\alpha'}-\vb*{x}_{n})}+\delta_{gg'},\\
&T_{(\alpha l)(\alpha' l')}=[(1-tG)^{-1}]_{(\alpha l)(\alpha' l')}t_{\alpha'l'},\\
&[1-tG]_{(\alpha l)(\alpha' l')}=\delta_{\alpha\alpha'}\delta_{ll'}- t_{\alpha l}G_{(\alpha l)(\alpha' l')}, \\
&G_{(\alpha l)(\alpha' l')}=\sum_{n\in{\bm Z}}{}' {\rm e}^{{\rm i}k_xan}H_{l'-l}^{(1)}(q|\tilde{\bm x}_{\alpha}-\tilde{\bm x}_{\alpha'}-na\hat{x}|){\rm e}^{{\rm i}(l'-l)\phi(\tilde{\bm x}_{\alpha}-\tilde{\bm x}_{\alpha'}-na\hat{x})}. \label{Eq_1dlatsum}
\end{align}
\end{widetext}
Here, index $\alpha(\alpha')$ is for cylinders per 1D unit cell of lattice constant $a$, index $l(l')$ refers to the 2D angular momentum, $\tilde{\bm x}_\alpha$ is the center coordinate of the $\alpha$-th cylinder, $t_{\alpha l}$ is the so-called $t$ matrix (or the Mie-scattering coefficient) of the isolated $\alpha$-th cylinder \cite{Ochiai:S::65:p245111:2002}, $\phi({\bm K})$ is the azimuthal angle of 2D vector ${\bm K}$, and $H_{l}^{(1)}$ is the Hankel function of the first kind and integer order $l$. 
The prime in the lattice sum of $G_{(\alpha l)(\alpha' l')}$ represents that $n=0$ is excluded if $\alpha=\alpha'$. 
This lattice sum 
can be calculated numerically either directly for large $\Im[q]$ or by the Ewald 
technique \cite{Ohtaka:U:A::57:p2550-2568:1998}.

In this case, the number $M$ of input channels is equal to the number of reciprocal lattices $g(g')$ taken into account in the numerical calculation.      

In what follows, we employ this numerical $S$ matrix for various calculations.

\section{non-Hermitian skin effect}
 
Let us consider a 2D PhC composed of dielectric cylinders with an optical gain.  The optical gain is represented by a negative imaginary part in the dielectric constant of the cylinders. 
This non-Hermiticity of the optical gain corresponds to a complex on-site (or, in other words, scalar) potential in a tight-binding picture   
of the system, instead of the complex vector potential of the Hatano-Nelson model \cite{PhysRevLett.77.570}. However, we should remind the reader that the tight-binding picture is available for limited cases, e.g.,   coupled cavity arrays, in photonic systems.

Let us further assume a composite square-lattice PhC with two cylinders per unit cell.  The positions of the cylinders are taken to be $(-a/8,-a/8)$ and $(a/8,a/8)$, where $a$ is the lattice constant. In this case, the system breaks the $x$ and $y$ inversion symmetries, whereas the 
exchange symmetry between $x$ and $y$ holds. 
In addition, the reciprocity, namely, the symmetry under  the transpose of the permittivity tensor, results in  $\omega(-k_x,-k_y)=\omega(k_x,k_y)$, where $\omega(k_x,k_y)$ is the complex eigenfrequency  under Bloch momentum $(k_x,k_y)$ \cite{PhysRevB.104.125416}.

These symmetry properties imply that the NHSE and the point-(pseudo)gap topology \cite{PhysRevX.8.031079} emerge in the boundary parallel or perpendicular to the $\Gamma$X direction, whereas they are forbidden in the boundary parallel or perpendicular to the $\Gamma$M direction.  
The breaking $y$ inversion symmetry results in $\omega(k_x,-k_y)\ne\omega(k_x,k_y)$, however at $k_x=0,\pm \pi/a$, the equality is recovered by the reciprocity.  Therefore, if we fix $k_x(\ne 0,\pm \pi/a)$, we have a loop in the complex frequency plane as we scan 
$k_y$ from $-\pi/a$ to $\pi/a$, giving rise to a point-(pseudo)gap and the NHSE. 
In contrast, the loop does not emerge in the $\Gamma$M direction. Consequently, the NHSE does not emerge.  In the boundary parallel to $\Gamma$M, the exchange symmetry directly results in  $\omega(k_\|,-k_\perp)=\omega(k_\|,k_\perp)$. 
In the boundary perpendicular to $\Gamma$M, the exchange symmetry gives  $\omega(-k_\|,k_\perp)=\omega(k_\|,k_\perp)$. Combining with the reciprocity $\omega(-k_\|,-k_\perp)=\omega(k_\|,k_\perp)$, we have 
again $\omega(k_\|,-k_\perp)=\omega(k_\|,k_\perp)$. 
Therefore, if we fix $k_\|$ and scan $k_\perp$, the point-(pseudo)gap loop is forbidden.

\begin{figure}
\includegraphics[width=0.45\textwidth]{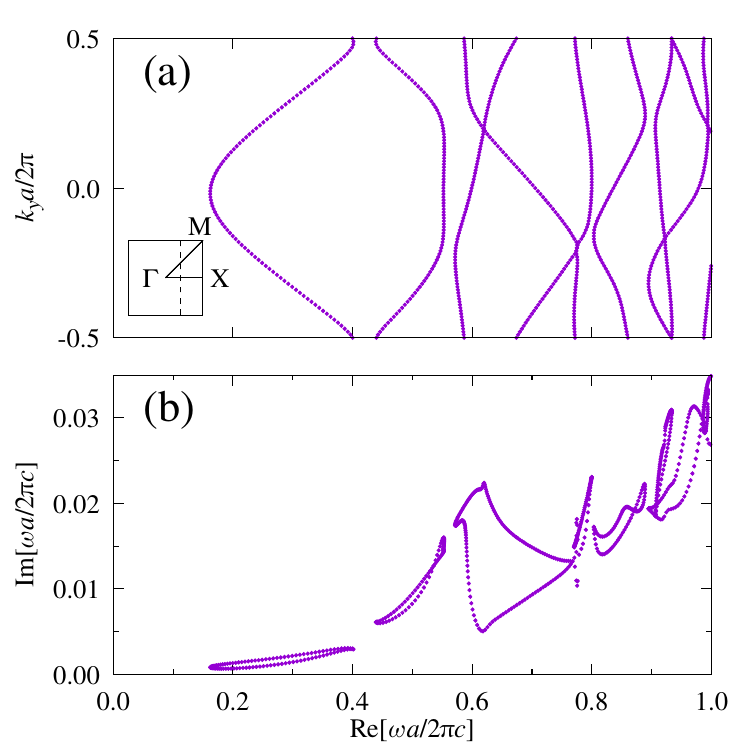} 
\caption{\label{Fig_bd}The photonic band structure of the TE polarization in the composite square lattice of dielectric cylinders with gain. 
The band structure is plotted (a) on the plane of the real part of the angular frequency $\Re[\omega]$ and (real) momentum $k_y$  and (b) on the complex frequency plane.  The cylinder has dielectric constant $\epsilon=12-\ii$ and radius $r=\sqrt{2}a/8$, where $a$ is the lattice constant of the square lattice.  
The cylinders are placed at $(-a/8,-a/8)$ and $(a/8,a/8)$ in a unit cell. The Bloch momentum $k_x$ is kept fixed at $0.4\pi/a$.  The inset in (a) shows the first Brillouin zone and the dashed line represents the scanned momentum axis.  	}
\end{figure}

Figure \ref{Fig_bd} shows the photonic band structure of the system with gain.  The band structure is of the TE polarization and was calculated by the PWE with the Ho-Chan-Soukoulis method \cite{ho1990epg}.
The $y$-inversion symmetry is broken as shown in Fig. \ref{Fig_bd}(a), resulting in the multiple loops in 
Fig. \ref{Fig_bd}(b) and a point-(pseudo)gap topology. 
A large loop is found around $\Re[\omega a/2\pi c]= 0.7$. This loop is not caused by a single band, but 
by the three bands entangled in the 
$(k_y,\Re[\omega],\Im[\omega])$ space as seen in Fig. \ref{Fig_bd}(a). 
In contrast, the loops around $\Re[\omega a/2\pi c]= 0.3$ and 0.5 consist of the respective single band.

\begin{figure*}
	\includegraphics[width=0.9\textwidth]{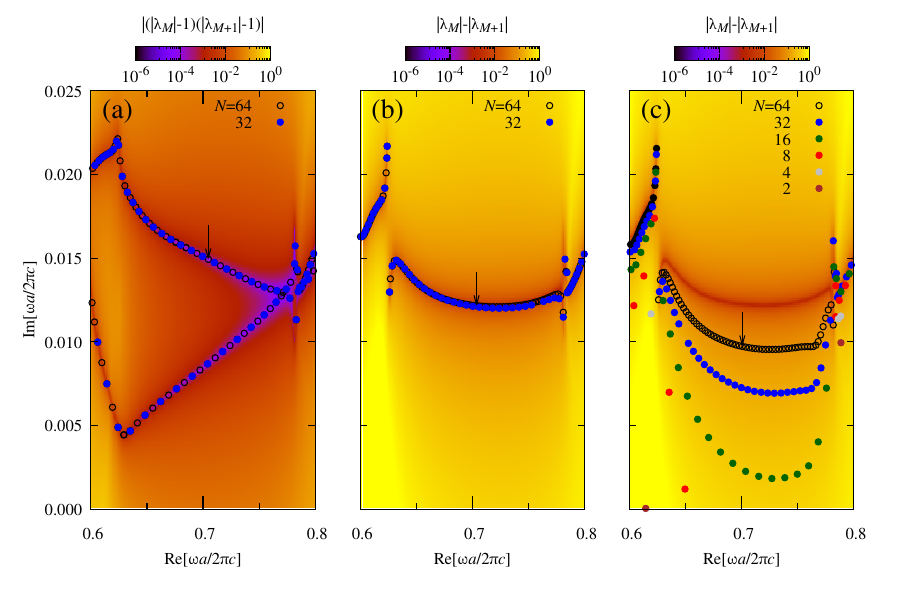}	
	\caption{\label{Fig_bd2} (a) The Bloch-band map 
 overlaid by the spectrum of finite-$N$ eigenmodes under the periodic boundary condition.  (b) The non-Bloch-band map 
 overlaid by the spectrum of finite-$N$ eigenmodes under the perfect- electric-conductor boundary condition. 
 (c) The non-Bloch-band map 
 overlaid by the spectrum of finite-$N$ eigenmodes under the absorbing open-boundary condition.   
  The Bloch-band map is the contour plot of $|(|\lambda_M|-1)(|\lambda_{M+1}|-1)|$ whose zeros correspond to the Bloch-band dispersion. 
 The non-Bloch-band map is the contour plot of $|\lambda_M|-|\lambda_{M+1}|$ whose zeros correspond to the non-Bloch-band dispersion. 	}
\end{figure*}

Figure \ref{Fig_bd2} shows the Bloch-band and non-Bloch-band maps overlaid by the complex  eigenfrequency spectra under the periodic, PEC (fixed-end open), and absorbing boundary conditions of finite 
$N$, calculated with the layer KKR method. 
The Bloch-band map is the contour plot of 
$|(|\lambda_M|-1)(|\lambda_{M+1}|-1)|$, whose zeros form curves in the complex frequency plane. These curves correspond to the Bloch-band dispersion and coincide with those in Fig. \ref{Fig_bd}(b).  
The non-Bloch-band map is the contour plot of 
 $|\lambda_M|-|\lambda_{M+1}|$, whose zeros form curves of the non-Bloch-band dispersion. 
We can see that these curves are completely different between Figs. \ref{Fig_bd2}(a) and \ref{Fig_bd2}(b).  
We also see that the finite-$N$ eigenmodes under the periodic boundary condition follow the Bloch-band dispersion, whereas those under the PEC boundary condition follow the non-Bloch-band dispersion. 
This property indicates that the NHSE occurs for the eigenmodes under the PEC boundary condition.

Remarkably, a strong $N$ dependence is observed for the absorbing open-boundary modes.  They tend to converge to the non-Bloch-band dispersion at $N\to\infty$. 
In contrast, the eigenmodes under the PEC boundary condition 
converge rapidly to the non-Bloch-band dispersion.

\begin{figure}
	\includegraphics[width=0.45\textwidth]{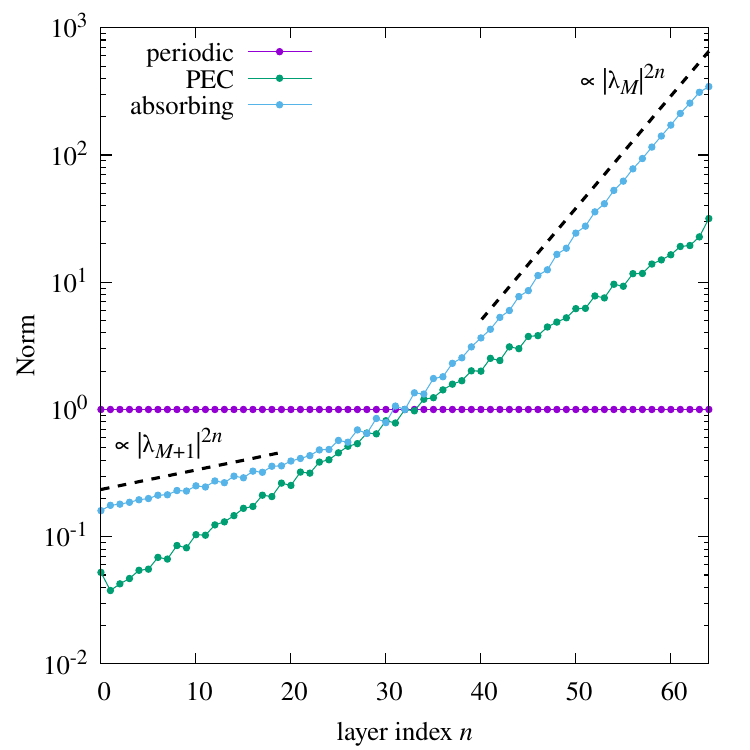}	
	\caption{\label{Fig_nhse} The norm ${\cal N}_n$ of the plane-wave expansion coefficients of the finite-$N$ eigenmodes under the periodic, PEC, and absorbing boundary condition, as a function of layer index $n$.  The total number of the layers is taken to be $N=64$. The eigenmodes are indicated by the arrows in Fig. \ref{Fig_bd2}. 	The dashed lines represent the exponential curves of $|\lambda_M|^{2n}$ and $|\lambda_{M+1}|^{2n}$ at the eigenfrequency of the absorbing boundary mode.}
\end{figure}

Figure \ref{Fig_nhse} shows a comparison of the eigenmodes around $\Re[\omega a/2\pi c]=0.7$, regarding the field profiles.    
Here we plot the norm ${\cal N}_n$ of the PWE coefficients in the empty region between the $n$-th and $(n+1)$-th layer, as a function of $n$. 
The norm is defined as  
\begin{align}
{\cal N}_n=\sum_g ( |a_{ng}^+|^2 + |a_{ng}^-|^2),	
\end{align}
where we take the normalization of ${\cal N}_{N/2}=1$. 
We can see that the eigenmodes under the PEC and absorbing boundaries are localized near the top ($n=N$) boundary. The PEC mode behaves like a single exponential, and the absorbing boundary mode behaves like a sum of two exponential terms.
In contrast, the eigenmode under the periodic boundary condition exhibits the constant norm.

The composite spatial decay in the NHSE of the absorbing boundary mode is a general feature reflecting the slow convergence to the non-Bloch-band dispersion at finite $N$.  
The spatial decays of the skin modes are described by the two eigenvalues $\lambda_M$ and $\lambda_{M+1}$ of the transfer matrix, located near the unit circle.  The other eigenvalues are not relevant.  The two exponential terms of the absorbing boundary mode are well approximated by $|\lambda_M|^{2n}$  and $|\lambda_{M+1}|^{2n}$ as shown in Fig. \ref{Fig_nhse}.  
The PEC mode of $N=64$ is found on the non-Bloch-band dispersion defined by $|\lambda_M|=|\lambda_{M+1}|$, so that these two terms are almost the same, showing the single exponential decay.   
The absorbing boundary mode is found off the non-Bloch-band dispersion even at $N=64$, so that $|\lambda_M|\ne |\lambda_{M+1}|$, showing the two exponential terms.

If we invert $k_x$, the PEC and absorbing boundary modes of Fig. \ref{Fig_nhse} are found at the same complex eigenfrequencies but localized near the bottom ($n=1$) layer.  
We also note that, if the non-Hermiticity is introduced as the loss by the complex conjugation of the dielectric constant, 
the band structure in Fig. \ref{Fig_bd} is flipped to negative $\Im[\omega]$ regions. Accordingly, 
the field localization of the NHSE of Fig. \ref{Fig_nhse} is switched to the bottom layer.

\section{Lasing}

The absorbing open-boundary modes correspond to the poles of the $S$ matrix in the complex frequency plane. The $S$ matrix defines the linear relation between the input and output coefficients. Therefore,  if a pole is on the real frequency axis, it represents a finite output under a vanishing input of the real frequency.  This is simply the lasing condition. 
Thus, the absorbing open-boundary modes are related to the lasing.

Even if the pole is close to but not on the real axis, it strongly affects the amplification of the incident light of a real frequency.  As shown in Fig. \ref{Fig_bd2} (c), the distribution of the poles depends on the number of PhC layers. 
The poles become dilute and closer to the real axis with reducing $N$. 
There is a trade-off between the rates and channels of the amplification.  
That is, the large amplification is obtained for smaller $N$ as the pole becomes closer to the real axis. 
However, the chance of the amplification is limited in a given interval of frequency as the poles become dilute.

The above trend also suggests that there is an optimal gain. If we reduce the gain, the eigenmodes become closer to the real axis. Thus, we have many chances of the lasing. However, smaller gains limit the rate of the amplification.

\begin{figure}
	\includegraphics[width=0.45\textwidth]{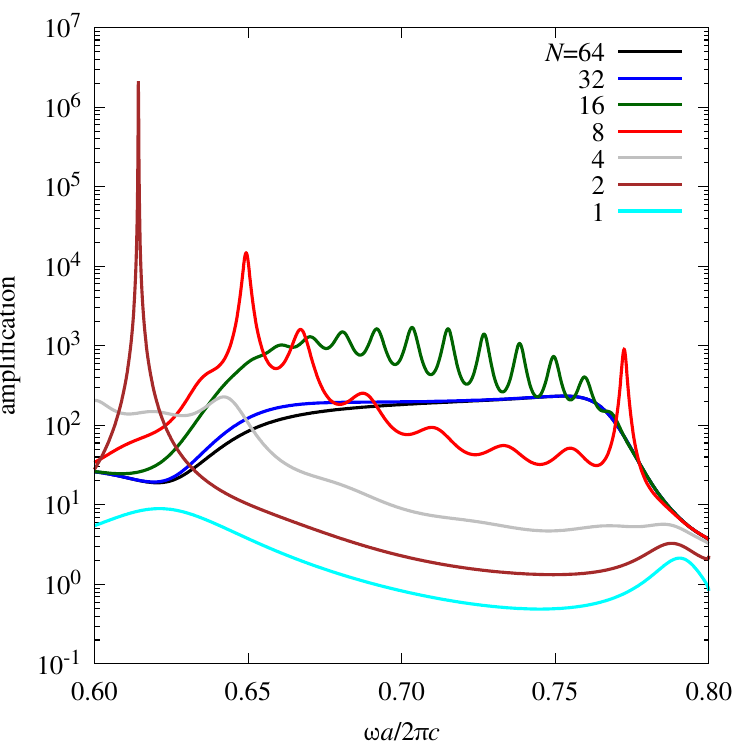}	
	\caption{\label{Fig_amp} The amplification ${\cal A}_N$ spectra of the composite square-lattice PhC with various $N$. The incident plane-wave light has the real angular frequency $\omega$ and real parallel momentum $k_x(=0.4\pi/a)$. 	}
\end{figure}

Figure \ref{Fig_amp} shows the amplification spectrum under the plane wave incidence.
Here, the incident light is coming from the bottom (of Fig. \ref{Fig_geoPhC}) with a real frequency. The PWE coefficients of the incident light are $a_{0g}^+=\delta_{g0}$ and $a_{Ng}^-=0$. 
The rate of the amplification ${\cal A}_N$ is defined by the minus  absorption, namely, 
\begin{align}
{\cal A}_N = -1+\sum_{g\in\textrm{open}}\frac{\Gamma_g}{\Gamma_0}(|a_{Ng}^+|^2+|a_{0g}^-|^2). 
\end{align}
It must vanish by energy conservation if there is no gain and loss. 
The amplification is strongly enhanced at $N=2$ around $\omega a/2\pi c=0.614$ and at $N=8$ around $\omega a/2\pi c=0.649$. A sequence of peaks is also found at $N=8$ and 16.  
However, no marked peaks are found at $N=32$ and 64, showing the 
saturation of the amplification spectra with increasing $N$.

The shift of the peak frequencies as a function of $N$ can be understood as follows.  At a particular $N$, we have a sequence of the poles of $S_N$. As we change $N$, the number of poles and their positions change. Among the poles, the closest one to the real axis of frequency strongly affects the amplification rate. As a result, the peak position changes with $N$. 
Also, $\Im[\omega]$ of the poles increases and converges to the non-Bloch-band dispersion with increasing $N$, so that the saturation of the amplification occurs.

In addition, the imaginary part in the eigenfrequency generally increases with increasing gain, so that the amplification tends to saturate with increasing gain.

We also note that when we scan a wider frequency range, a general trend of enhanced amplification near the band edges  \cite{Sakoda:O:U::4:pU1-U9:1999} is observed.

\begin{figure*}
	\includegraphics[width=0.9\textwidth]{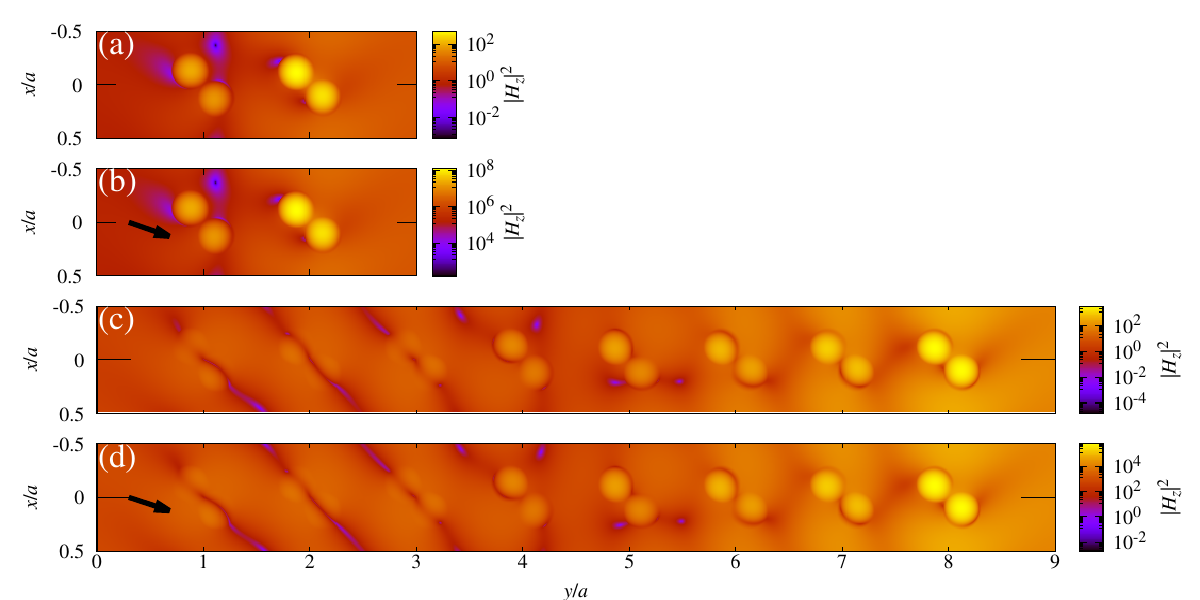}	
	\caption{\label{Fig_conf} The field profiles $|H_z|^2$  of (a) the absorbing open-boundary mode of $N=2$ at $\omega a/2\pi c\simeq 0.614+0.4\times 10^{-4}\ii$, (b) the near-lasing configuration of $N=2$ at $\omega a/2\pi c=0.614$ under the plane-wave incidence, (c) the absorbing open-boundary mode of $N=8$ at $\omega a/2\pi c\simeq 0.649+0.12\times 10^{-2}\ii$, and (d) the near-lasing configuration of $N=8$ at $\omega a/2\pi c=0.649$ under the plane-wave incidence. In (a) and (c), the fields are normalized such that $\sum_g |a_{N/2\;g}^+|^2=1$.  In (b) and (d), the incident light $H_z^0$ from the left has the unit amplitude $H_z^0=\exp(\ii (k_xx + \Gamma_0y))$. The arrows in (b) and (d) represent the wave vectors of the incident plane waves. 
	}
\end{figure*}

Figure \ref{Fig_conf} shows the field profiles
of the absorbing open-boundary modes nearest and next-nearest to the real frequency axis, and of the (almost) lasing configurations under the incident plane-wave light at the corresponding peak frequencies of Fig. \ref{Fig_amp}.  
The field patterns in Figs. \ref{Fig_conf}(a) and \ref{Fig_conf}(b)  [or \ref{Fig_conf}(c) and \ref{Fig_conf}(d)] resemble each other very closely, except for the intensity due to the normalization scheme.   
This resemblance indicates that the amplification is caused by the corresponding  absorbing open-boundary mode. 
As the mode is localized near the top (right) boundary in \ref{Fig_conf}(c), the amplification of the incident light is forward oriented in \ref{Fig_conf}(d).

Similarly, we can show that the Fabri-Perot-like fringes of the amplification spectrum of $N=16$ in Fig. \ref{Fig_amp} are caused by the corresponding absorbing open-boundary modes.

\section{Conclusion}
In summary, we have explored the absorbing open-boundary modes in a non-Hermitian photonic crystal. They have continuum spectra of the non-Bloch-band dispersion in the infinite-system-size limit but exhibit a substantial  deviation from the limiting curve at finite system sizes. They show the NHSE as the conventional fixed-end open-boundary modes, if the point-(pseudo)gap topology is available. If the absorbing open-boundary modes emerge in the vicinity of the real frequency axis, they work as lasing modes.





\begin{acknowledgments}
This work was partially supported by JSPS KAKENHI Grant No. 22K03488. 
\end{acknowledgments}

%

\end{document}